# Disproof of solar influence on the decay rates of $^{90}$Sr/$^{90}$Y

Karsten Kossert[*] and Ole J. Nähle

*Physikalisch-Technische Bundesanstalt (PTB), Bundesallee 100, 38116 Braunschweig, Germany*

**Abstract**

A custom-built liquid scintillation counter was used for long-term measurements of $^{90}$Sr/$^{90}$Y sources. The detector system is equipped with an automated sample changer and three photomultiplier tubes, which makes the application of the triple-to-double coincidence ratio (TDCR) method possible. After decay correction, the measured decay rates were found to be stable and no annual oscillation could be observed. Thus, the findings of this work are in strong contradiction to those of Parkhomov [1] who reported on annual oscillations when measuring $^{90}$Sr/$^{90}$Y with a Geiger-Müller counter. Sturrock et al. [2] carried out a more detailed analysis of the experimental data from Parkhomov and claimed to have found correlations between the decay rates and processes inside the Sun. These findings are questionable, since they are based on inappropriate experimental data as is demonstrated in this work. A frequency analysis of our activity data does not show any significant periodicity.

*Keywords:* Liquid scintillation counting; TDCR; $^{90}$Sr/$^{90}$Y; disproof of solar influence on the decay rate

## 1. Introduction

Recently, Parkhomov[1] published results on long-term measurements of $^{90}$Sr/$^{90}$Y which were obtained by means of a Geiger-Müller counter. The data show large fluctuations and Parkhomov identified annual and monthly oscillations. According to Parkhomov, the data support the contentious assertion of correlations between the radioactive decay and the Earth-Sun distance as reported in [3]. In a further article, Sturrock et al. report on correlations between the $^{90}$Sr/$^{90}$Y data and inner Sun effects [2].

The articles by Parkhomov [1] and Sturrock et al. [2] do not contain any detailed description about the source and its activity, the detector, the experimental conditions, the source-detector geometry or the duration of the measurements. Moreover, none of the articles presents a detailed uncertainty consideration. Thus, it is difficult to judge whether the data are suitable for a correlation analysis. The main fault in their works is obvious: They equate observed fluctuations in the instrument readings with fluctuations of decay rates. This is nonsense, however, since the fluctuating parameter is the instrument detection efficiency as we will also demonstrate with our data. Parkhomov excludes environmental parameters like the changes in temperature as a reason for the oscillation, since similar oscillations were observed in other laboratories when measuring other isotopes. However, this statement is not convincing, since a correlation is already given by the fact that winter is a common phenomenon which influences most laboratories in the northern hemisphere where the data were taken.

It is also known that gas counters like Geiger-Müller counters are very sensitive to changes in environmental parameters. In particular, when external solid sources are measured, the counting efficiency very much depends on parameters like air humidity, temperature and air pressure. Thus, we decided to carry out measurements by means of $4\pi$ liquid scintillation counting. In a previous study of $^{36}$Cl [4], the results were very robust and show no oscillations. Thus, the results show that other observations of oscillations (see, e.g., Ref. [5]) are due to ill-suited detectors rather than to solar influence or any other "new physics" [6].

---

[*] corresponding author, e-mail: karsten.kossert@ptb.de

In this work, we present new $^{90}$Sr/$^{90}$Y data which were measured with a new TDCR counter at PTB which is equipped with an automated sample changer. The data show neither an annual nor a monthly oscillation and, thus, the conclusions from Parkhomov [1] and Sturrock et al. [2] are found to be false.

## 2. TDCR measurements

Three samples were prepared in April 2013 using 15 mL Ultima Gold$^{TM}$ scintillation cocktail in 20 mL borosilicate glass vials with low potassium content. The scintillation cocktail is based on the solvent di-isopropylnaphthalene which causes slightly yellow colouring after some years. Thus, we have taken a new batch of this scintillation cocktail (expiry date July 2014) to avoid any colour quenching effects during the period of observation.

About 0.5 mL of distilled water and weighed portions of about 500 mg of a $^{90}$Sr/$^{90}$Y solution with an activity concentration of 3.86 kBq·g$^{-1}$ on the reference date of 1 January 2013 (0 h UTC) were added. The aqueous $^{90}$Sr solution contained SrCl$_2$·6H$_2$O and YCl$_3$·6H$_2$O as non-radioactive carriers with concentrations of about 50 mg·L$^{-1}$ and 46 mg·L$^{-1}$, respectively, in 0.1 M HCl. The solution was purchased in 2005 and, thus, $^{90}$Sr and $^{90}$Y are in secular equilibrium. An aliquot of the solution was measured by means of gamma-ray spectrometers and no photon-emitting impurity could be detected. The background counting rates were measured with a blank sample which was prepared with 15 mL Ultima Gold and 1 mL of distilled water. Small amounts of nitromethane (CH$_3$NO$_2$) were added to two samples in order to vary the counting efficiency. In the following, the background sample is denoted as sample No. 1, whereas the three samples containing $^{90}$Sr/$^{90}$Y are labelled 2, 3 and 4, respectively.

The samples were measured in a new custom-built TDCR detector with an automated sample changer. The counter is an improved version of the first TDCR counter at PTB which was described in Ref. [7]. The temperature in the measurement room was found to be in the range from 18°C to 22°C during most of the TDCR measurements. However, due to several periods of maintenance the temperature was sometimes out of the stated range with a minimum temperature of 16.7°C and a maximum temperature of 29.7°C during individual days. The TDCR system consists of an optical chamber with three photomultiplier tubes (PMTs) surrounding a liquid scintillation sample in the centre. For this work, three Hamamatsu R331-05 PMTs were mounted. The optical chamber is made of the diffuse reflecting material OP.DI.MA (ODM98) produced by Gigahertz Optik GmbH, with a reflectivity of more than 98 % over a wide wavelength range. The high reflectivity is important to achieve a high counting efficiency to minimize environmental influence and to reduce the uncertainty of the determined decay rates.

The anode signals of the PMTs are amplified by a CAEN N978 fast amplifier and discriminated by an Ortec 935 Constant-Fraction Discriminator. The discrimination threshold was adjusted just below the single electron peak analysing the pulse height spectra of each photomultiplier with an analog to digital converter. The threshold was checked regularly at least three times a year and no drift of the threshold could be observed. The digital pulses from the discriminator are fed into an FPGA-based module developed at PTB [8]. The module is referred to as the 4KAM system and emulates the behaviour of the analog MAC3 module [9]. The system is a lifetime-based coincidence module with an extending dead time to suppress afterpulses.

The net counting rates for triple and double coincidences $R_T$ and $R_D$ from the TDCR system are used to determine the experimental TDCR value, i.e. TDCR $= R_T/R_D$. The corresponding counting efficiencies for triple and double coincidences $\varepsilon_T$ and $\varepsilon_D$ were computed by means of the MICELLE2 program [10] using nuclear decay data from Ref. [11]. The computation results were used to get two functions $\varepsilon_T(\text{TDCR})$ and $\varepsilon_D(\text{TDCR})$. Consequently, the counting efficiencies can be obtained from the experimentally determined TDCR value. Details about the TDCR method are described in Ref. [12].

Several measurements were carried out in the period between 24 April 2013 and 26 May 2014. The minimum measurement duration per single measurement of an individual sample is 900 s, but many measurements were carried out using 1200 s, 1800 s and 3600 s, respectively.

## 3. Analysis and results

Figure 1 shows the net counting rates for triple coincidences $R_T$ vs. time for sample 2. The counting rates were corrected for decay using the $^{90}$Sr half-life $T_{1/2} = 10519(26)$ d [11] and then normalized to a mean value of 1. The data show a lower spread than those data from Ref. [1]. Nevertheless, Fig. 1 reveals some variations with time – in particular a slight decreasing trend. Here, we emphasize again that Fig. 1 does **not** show variations of decay rates with time, but variations of the instrument readings.

The instrument reading (counting rate for triple coincidences $R_T$) is given by

$$R_T = A \cdot \varepsilon_T,\qquad(1)$$

where $A$ is the sample's activity and $\varepsilon_T$ the counting efficiency which might vary due to changes in the experimental conditions. The counting rate for double coincidences $R_D$ can be expressed in a similar manner.

Fig. 2 shows the counting efficiency $\varepsilon_T$ as a function of time normalized to a mean value of 1. The counting efficiency was derived theoretically from the measured TDCR value and the function $\varepsilon_T(\text{TDCR})$. The plot shows a similar decreasing trend and timely variations. It should be emphasized that $\varepsilon_T(\text{TDCR})$ is not calculated from the known activity $A$ and the measured counting rates but rather only from the TDCR value.

Finally, Fig. 3 shows the normalized activity $A$ which was derived from the triple coincidence counting rate and the corresponding counting efficiency according to

$$A = R_T / \varepsilon_T(\text{TDCR}).\qquad(2)$$

No significant trend can be seen in Fig. 3 and the data also give no hint of any oscillation. It should be noted that the spread of the data changes with time can be well explained by the fact that different counting times were adjusted (see above).

The analysis shows that there are no oscillations in the decay rate. It also makes the major faults from Sturrock et al. clear: They interpreted variations in the counting rate (left-hand side of Eq. 1) as variations in the decay rate (left-hand side of Eq. 2) and ignored the fact that variations are caused by variations in the counting efficiency. The TDCR method allows a calculation of the counting efficiencies based on a theoretical model and, thus, proves that indeed we observe a change in counting efficiency rather than a change in decay rates. This shows the drawback of relative measurements which give no direct information about the instrument detection efficiency. This also holds for simple Geiger-Müller counters as well as for ionization chambers and gamma-ray spectrometers. Thus, we propose the usage of primary methods for the study of potential variations in decay rates. Primary methods yield information on the efficiency of the source-detector system and the decay rate (activity).

The determined activities of samples 3 and 4 are shown in Figures 4 and 5, respectively. Again, the activities were obtained from the triple-coincidence counting rate and the corresponding counting efficiencies $\varepsilon_T$(TDCR). Very similar results were obtained when the activities were calculated from the double-coincidence counting rates and the efficiency $\varepsilon_D$(TDCR).

The data shown in Figures 3, 4 and 5 were also used for spectrum analyses of unevenly sampled data using the Lomb-Scargle procedure [13-15]. In some cases, the power of certain frequencies passed the significance level $\alpha = 0.5$, but the power was lower at the same frequency for at least one other sample. Thus, none of the identified frequencies can be considered as being significant. The periodograms are shown in Figures 6, 7 and 8. It is important to note that none of the spectra shows a pronounced peak at 1 $y^{-1}$.There is also no significant peak at about 12 $y^{-1}$ and, hence, an influence of the Moon, as proposed by Parkhomov [1], can be excluded as well. We extended the search towards higher frequencies, but could not identify any significant frequency common to all three samples.

Fig. 9 shows the results obtained in this work compared to the results from Sturrock et al. [2]. The data of this work correspond to the mean value of all three $^{90}$Sr/$^{90}$Y samples comprising 20 single measurements per sample. As a result, the relative statistical uncertainty $N^{-1/2}$, estimated from the total number of counts $N$, is lower than $1\times10^{-4}$. An uncertainty budget in compliance to [16] for the normalized decay rate is shown in Table 1. The scatter of the normalized decay rates from the LS measurements in Fig. 9 is much lower than the spread of the counting rates (not decay rates!) published by Sturrock et al. This is no surprise, since gas counters such as the used Geiger-Müller counter are known to be sensitive to environmental parameters.

The scatter of the mean decay rates determined at PTB is in the order of $\pm3\times10^{-4}$ and, consequently, they are in accordance with the estimated uncertainty (Table 1).

Fig. 9 also shows the squared inverse Sun-Earth distance as a solid line. The curve is adjusted using the dates for perihelion and aphelion from the United States Naval Observatory [17]. Obviously, the oscillation agrees neither in amplitude nor in phase with the variation seen in the normalized counting rates as published by Sturrock et al. [2] and, consequently, their results contradict their own previous assertion which claims a correlation between radioactive decay rates and the Sun-Earth distance [3].

## 4. Summary and conclusion

It was demonstrated that primary activity standardization techniques are required to study potential variations in decay rates, since other methods cannot provide direct information about the detection efficiency of the source-detector system. In most detector systems, the detection efficiency varies with time – as it does for our TDCR system which is even more stable than gas counters.

The TDCR method yields clear information on the counting efficiency and the decay rates. For $^{90}$Sr/$^{90}$Y no periodicity could be found when analysing the decay rate. Thus, the assertions made in Refs. [1], [2] and [18] are false and there is no need for "new physics" [6]. The suggestion "that neutrinos interact with nuclei via a new potential", which some researchers consider to be "a manifestation of a new boson" [18], is ridiculous.

**Figures**

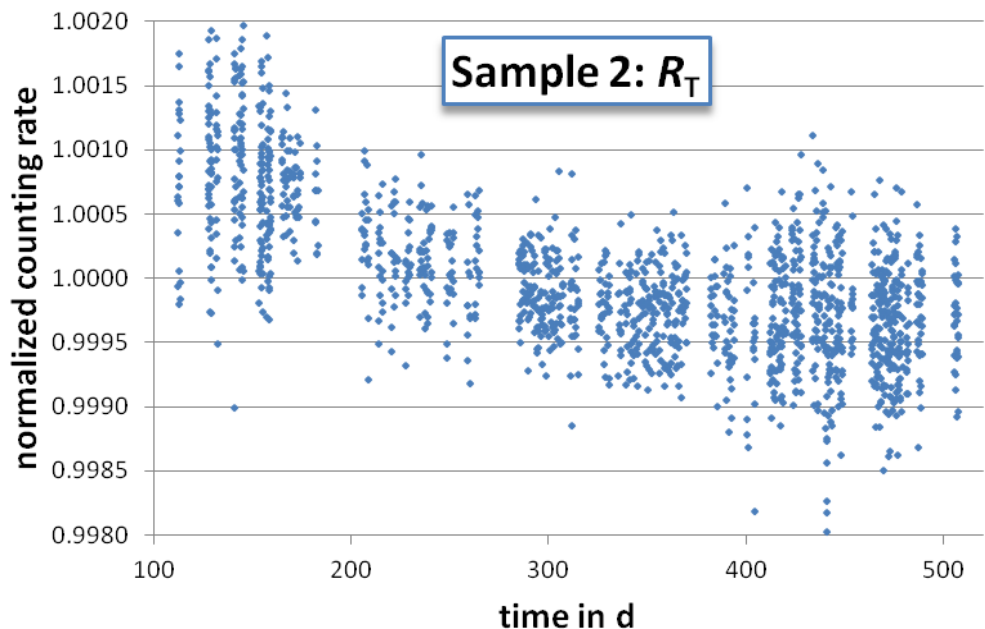

Figure 1

Net counting rates $R_T$ of sample 2. The counting rates were corrected for decay and then normalized to a mean value of 1. The time is shown in days since 1 January 2013.

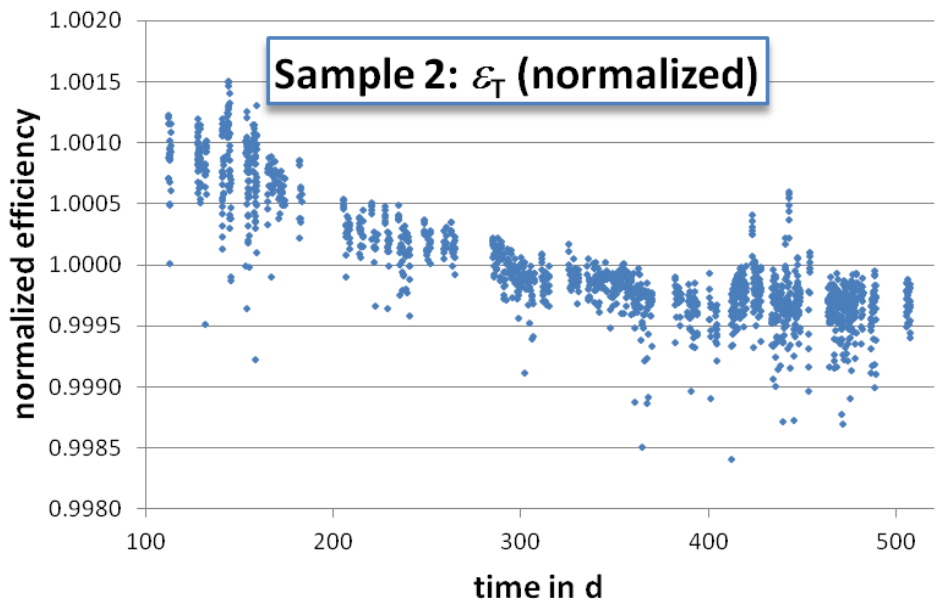

Figure 2
Normalized counting efficiency $\varepsilon_\text{T}$ of sample 2. The time is shown in days since 1 January 2013.

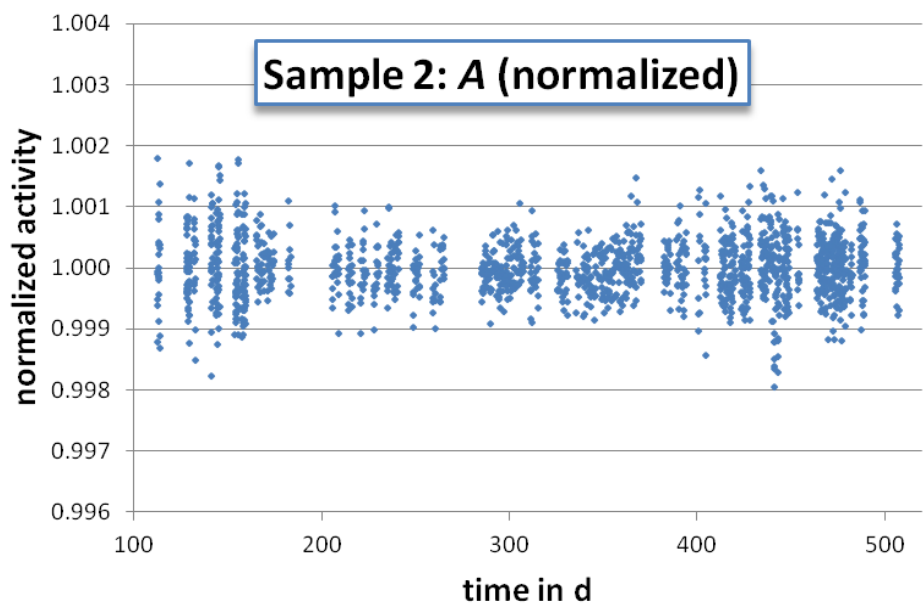

Figure 3
Normalized activity derived from triple coincidences measured with sample 2 and theoretically calculated counting efficiencies. The time is shown in days since 1 January 2013.

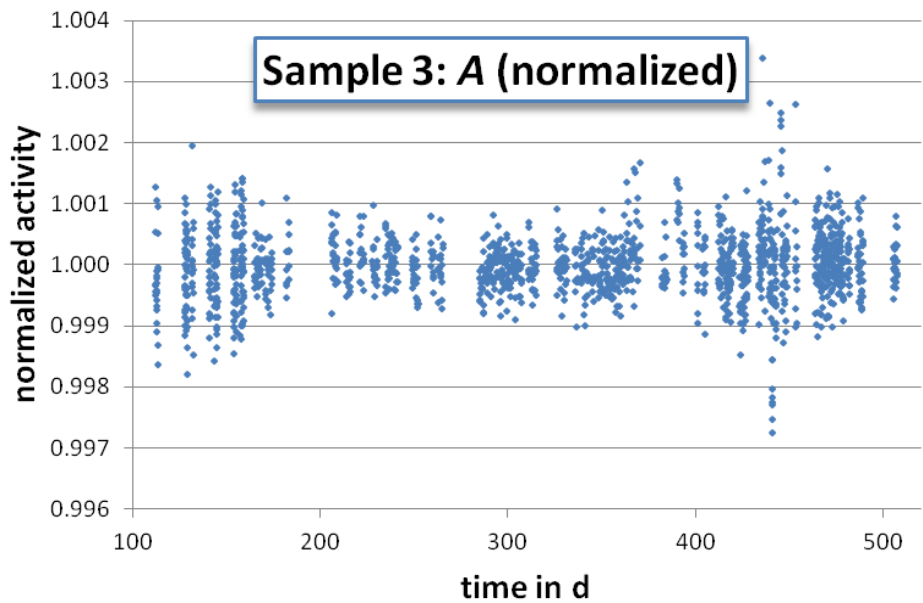

Figure 4
Normalized activity derived from triple coincidences measured with sample 3. The time is
shown in days since 1 January 2013.

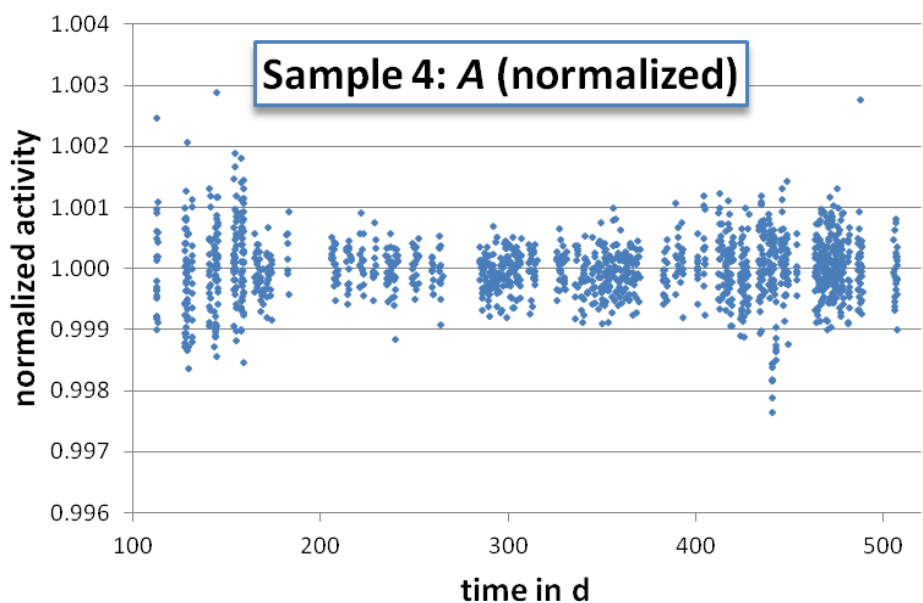

Figure 5
Normalized activity derived from triple coincidences measured with sample 4. The time is
shown in days since 1 January 2013.

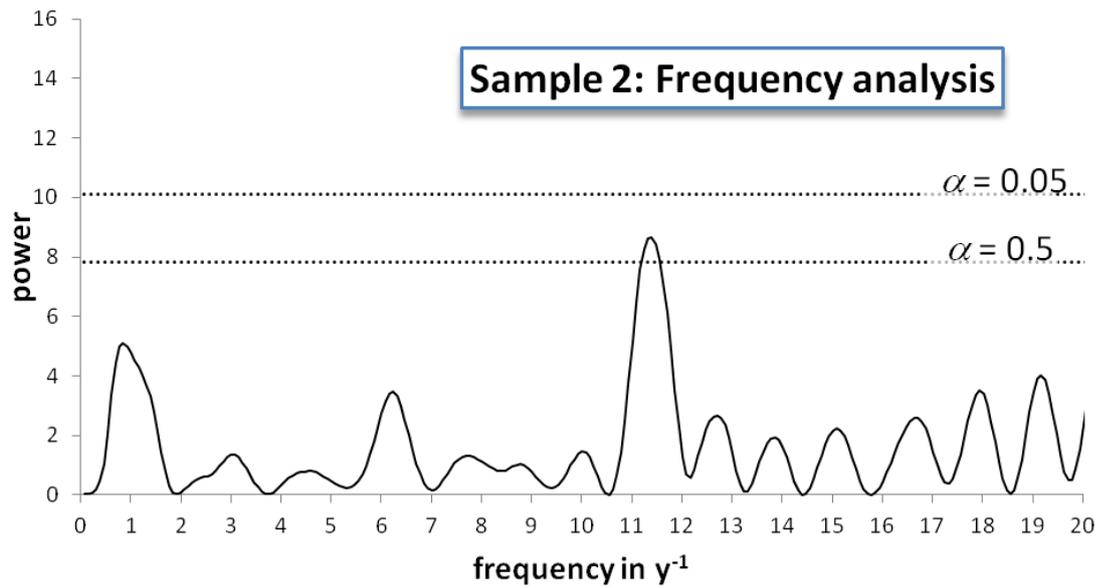

Figure 6
Power spectrum from the frequency analysis of the activity-vs.-time data (Fig. 3) of sample 2. The horizontal lines indicate the power for significance level of 0.5 and 0.05, respectively. The maximum is seen at about 11.5 $y^{-1}$, but the corresponding peak is missing in the plots of samples 3 and 4.

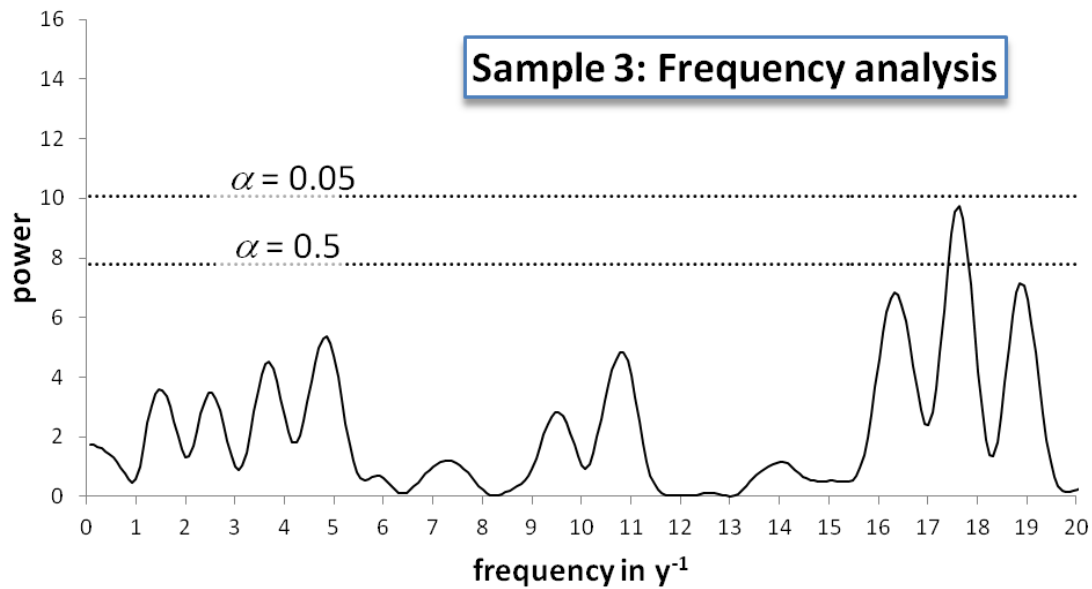

Figure 7
Power spectrum from the frequency analysis of the activity-vs.-time data (Fig. 4) of sample 3 (see also caption of Fig. 6).

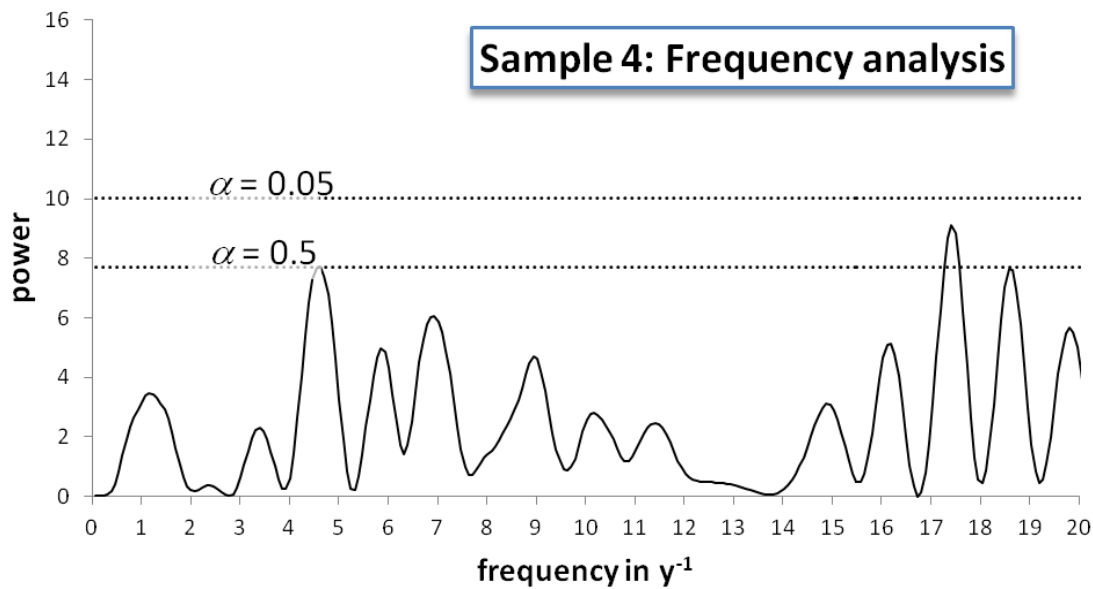

Figure 8
Power spectrum from the frequency analysis of the activity-vs.-time data (Fig. 5) of sample 4
(see also caption of Fig. 6).

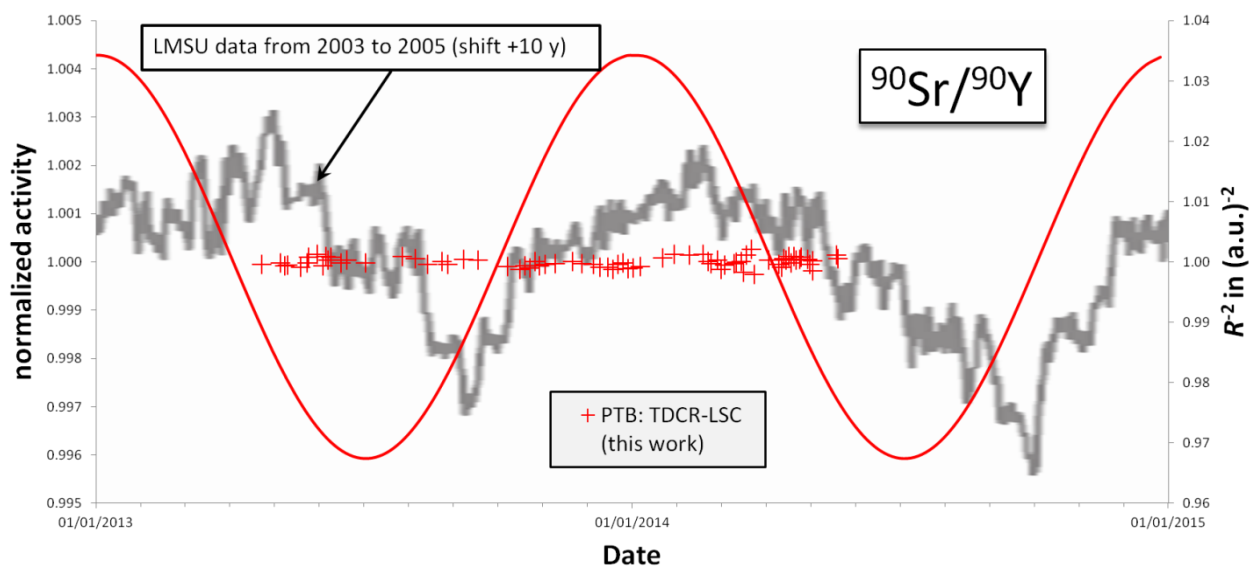

Figure 9
Normalized $^{90}$Sr/$^{90}$Y activities measured by means of TDCR in this work (+ symbols)
compared to normalized counting rates measured at LMSU as published by Sturrock et al. [2].
The LMSU data were taken from the years 2003 to 2005 and shifted to match the observation
interval of this work. The solid line represents the squared inverse Sun-Earth distance (right
ordinate).

Table 1

Standard uncertainty components of the normalized decay rate of the $^{90}$Sr/$^{90}$Y solution determined by means of TDCR.

| Component | Relative uncertainty component in % |
|---|:---:|
| Statistical uncertainty (see text) | 0.01 |
| Background | 0.02 |
| Time of measurements (starting time and duration (lifetime)) | 0.01 |
| TDCR value and fit | 0.01 |
| Decay correction | <0.01 |
| **Square root of the sum of quadratic components** | **0.03** |